\documentclass[12pt]{article}

\usepackage{RRA4}
\usepackage{epic}
\usepackage{eepic}
\usepackage{graphics}
\usepackage{index}
\usepackage[english]{babel}
\usepackage{latexsym}
\usepackage{amsthm}
\usepackage{color}

\usepackage{bbm}     
\usepackage{epsfig}   
\usepackage{amsmath,amssymb,graphicx}
\usepackage{epsfig}
\usepackage{epic,eepic}

\newcommand{\DS}[1]{{\displaystyle #1}}

\newcommand{\vhi}{\varphi}

\newcommand{\nbu}{\mathbbm{1}}

\newcommand{\p}{\mathbb{P}}
\newcommand{\e}{\mathbbm{E}}
\newcommand{\beq}{\begin{equation}}
\newcommand{\eeq}{\end{equation}}
\newcommand{\bd}{\begin{displaymath}}
\newcommand{\ed}{\end{displaymath}}
\renewcommand{\l}{\lambda}
\renewcommand{\geq}{\geqslant}
\renewcommand{\leq}{\leqslant}

\newtheorem{theorem}{{\bf Theorem}}
\newtheorem{lemm}[theorem]{{\bf Lemma}}

\renewcommand{\proof}{\noindent {\bf Proof. }}

\def\QED{\mbox{\rule[0pt]{1.5ex}{1.5ex}} \vspace{0.2cm}}
\def\cqfd{\hspace*{\fill}~\QED\par\endtrivlist\unskip}

\begin{document}

\RRdate{February 2008}
\RRNo{6443}
\RRauthor{Nathalie Mitton \thanks{INRIA Lille - Nord Europe/LIP6(USTL,CNRS), 
nathalie.mitton@inria.fr},
Katy Paroux \thanks{Universit\'e de Franche-Comt\'e,
katy.paroux@univ-fcomte.fr},
Bruno Sericola \thanks{INRIA Rennes - Bretagne Atlantique,
bruno.sericola@inria.fr},
S\'ebastien Tixeuil \thanks{INRIA Saclay - \^Ile-de-France/LIP6,
sebastien.tixeuil@lri.fr}}
\authorhead{N. Mitton, K. Paroux, B. Sericola \& S. Tixeuil} 
\RRetitle{Ascending runs in dependent uniformly distributed random variables:
Application to wireless networks}
\titlehead{Ascending runs in dependent uniformly distributed random variables}
\RRtitle{Sous-suites croissantes contigu\"es de variables al\'eatoires 
d\'ependantes uniform\'ement distribu\'ees: application aux r\'{e}seaux sans fil}
\RRabstract{    
We analyze in this paper the longest increasing contiguous sequence or
maximal ascending run of random variables with common uniform distribution
but not independent. Their dependence is characterized by the fact that two
successive random variables cannot take the same value.
Using a Markov chain approach, we study the distribution of the maximal
ascending run and we develop an algorithm to compute it.
This problem comes from the analysis of several self-organizing protocols
designed for large-scale wireless sensor networks,
and we show how our results apply to this domain.
}
\RRresume{Nous analysons dans cet article la plus longue sous-suite croissante
contigu\"e d'une suite de variables al\'eatoires de m\^eme distribution 
uniforme mais non 
ind\'ependantes. Leur d\'ependance est caract\'eris\'ee par le fait que
deux variables successives ne peuvent prendre la m\^eme valeur.
En utilisant une approche markovienne, nous \'etudions la distribution
de la plus longue sous-suite croissante contigu\"e et nous d\'eveloppons
un algorithme pour la calculer. Ce probl\`eme provient de l'analyse
de plusieurs protocoles auto-organisants pour les r\'eseaux de capteurs
sans fil \`{a} grande \'{e}chelle,
et nous montrons comment nos r\'esultats s'appliquent \`a ce domaine.
}

\RRmotcle{Cha\^{\i}nes de Markov, sous-suites croissantes contigu\"es, auto-stabilisation, temps de convergence.}
\RRkeyword{Markov chains, maximal ascending run, self-stabilization, convergence time.}
\RRprojets{Dionysos, Grand Large et Pops} 
\RRtheme{\THCom \THNum} 

\URRennes

\makeRR


\section{Introduction}
Let $X = (X_n)_{ n \geq 1}$ be a sequence of identically 
distributed random variables on the set $S = \{1,\ldots,m\}$.
As in \cite{Louchard02}, we define an ascending run as a contiguous
and increasing subsequence in the process $X$. For instance, with $m = 5$,
among the $20$ first following values of $X$: $23124342313451234341$, there are
$8$ ascending runs and the length of maximal ascending run is $4$.
More formally, an ascending run of length $\ell \geq 1$, starting at position 
$k \geq 1$,
is a subsequence
$(X_k,X_{k+1},\ldots,X_{k+\ell-1})$ such that
\begin{displaymath}X_{k-1} > X_k < X_{k+1}< \cdots < X_{k+\ell-1} > X_{k+\ell},
\end{displaymath}
where we set $X_0 = \infty$ in order to avoid special cases at the boundary.
Under the assumption that the distribution is discrete and the random 
variables are 
independent, several authors have studied the behaviour of the maximal ascending
run, as well as the longest  non-decreasing contiguous subsequence. 
The main results concern the asymptotic behaviour of these quantities when 
the number of 
random variables tends to infinity, see for example \cite{Frolov99} and 
\cite{Eryilmaz06} and the references therein.
Note that these two notions coincide when the common distribution is continuous.
In this case, the asymptotic behaviour is known and does not depend on the 
distribution, as shown in \cite{Frolov99}. 

We denote by $M_{n}$ the length of the
maximal ascending run among the first $n$ random variables.
The asymptotic behaviour of $M_{n}$ hardly depends on the common distribution 
of the random variables
$X_{k},\,k\geq 1$. Some results have been established for the geometric 
distribution
in \cite{Louchard03} where an equivalent of the law of $M_{n}$ is provided and
previously in \cite{Csaki96} where the almost-sure convergence is
studied, as well as for Poisson distribution.

In \cite{Louchard05}, the case of the uniform distribution on the set
$\{1,\ldots,s\}$ is investigated. The author considers the problem of the 
longest non-decreasing
contiguous subsequence and gives an equivalent of its law when $n$ is 
large and $s$ is fixed. 
The asymptotic equivalent of $\e(M_{n})$ is also conjectured.

In this paper, 
we consider a sequence $X = (X_n)_{ n \geq 1}$ of integer random variables
on the set $S = \{1, \ldots, m\}$, with $m \geq 2$.
The random variable $X_1$ is uniformly distributed on $S$ and, for $n \geq 2$,
$X_n$ is uniformly distributed on $S$ with the constraint $X_n \neq X_{n-1}$.
This process may be seen as the drawing of balls, numbered from $1$ to $m$  in
an urn where at each step the last ball drawn is kept outside the urn.
Thus we have, for every $i,j \in S$ and $n \geq 1$,
\begin{displaymath}\p(X_1 = i) = \frac{1}{m} \mbox{ and }
\p(X_n = j | X_{n-1} = i) = \frac{1_{\{i \neq j\}}}{m-1}.\end{displaymath}
By induction over $n$ and unconditioning, we get, for every $n \geq 1$ and
$i \in S$,
\begin{displaymath}\p(X_n = i) = \frac{1}{m}.\end{displaymath}
Hence the random variables $X_n$ are uniformly distributed on $S$ but  are 
not independent.
Using a Markov chain approach, we study the distribution of the maximal
ascending run and we develop an algorithm to compute it.
This problem comes from the analysis of self-organizing protocols
designed for large-scale wireless sensor networks,
and we show how our results apply to this domain.

The remainder of the paper is organized as follows.
In the next section, we use a Markov chain approach to study the behavior
of the sequence of ascending runs in the process $X$. In Section~\ref{sec:hitting}, we analyze
the hitting times of an ascending run of fixed length and we obtain the 
distribution of the maximal ascending $M_n$ over the $n$ first random variables
$X_1,\ldots,X_n$ using a Markov renewal argument. An algorithm to compute this 
distribution is developed in Section~\ref{sec:algorithm} and Section~\ref{sec:appli} is devoted to the 
practical implications of this work in large-scale wireless sensor networks.

\section{Associated Markov chain}
\label{sec:markov}

The process $X$ is obviously a Markov chain on  $S$.
As observed in \cite{Louchard03}, we can see the ascending runs as
a discrete-time process having two components: the value taken by the first 
element 
of the
ascending run and its length.
We denote this process by $Y = (V_k,L_k)_{ k \geq 1}$, where
$V_k$ is the value of the first element of the $k$th ascending run and 
$L_k$ is its length.
The state space of $Y$ is a subset $S^2$ we shall precise now.

Only the first ascending run
can start with the value $m$. Indeed, as soon as $k\geq 2$, the random 
variable $V_{k}$ takes its values in $\{1,\ldots,m-1\}$.
Moreover $V_{1}=X_{1}=m$ implies that $L_{1}=1$. Thus, for any $\ell\geq 2$, 
$(m,\ell)$ is not a state of $Y$ whereas $(m,1)$
is only an initial state that $Y$ will never visit again. 

We observe also that if $V_{k}=1$ then necessarily $L_{k}\geq 2$, which 
implies that ${(1,1)}$ is not a state of $Y$. 
Moreover $V_{k}=i$ implies that $L_{k}\leq m-i+1$. 

According to this behaviour, we have
$$Y_1 \in E \cup \{(m,1)\} \mbox{ and for $k \geq 2$, } Y_k \in E,$$
where
$$E=\{(i,\ell)\,|\, 1\leq i\leq m-1 \mbox{ and } 1\leq \ell \leq m-i+1\}
\setminus \{(1,1)\}.$$

\noindent We define the following useful quantities for any 
$i,j,\ell \in S$ and $k\geq 1$~:
\begin{eqnarray}
\Phi_\ell(i,j) & = & \p(V_{k+1} = j, L_k = \ell | V_k = i),\\
\vhi_\ell(i) & = & \p(L_{k} = \ell | V_k = i), \label{fiil} \\
\psi_\ell(i) & = & \p(L_{k} \geq \ell | V_k = i). \label{psiil}
\end{eqnarray}

\begin{theorem}\label{Ycdm}
The process $Y$ is a homogeneous Markov chain with transition probability
matrix $P$, which entries are given for 
any $(i,\ell) \in E \cup \{(m,1)\}$ and $(j,\l)\in E$ by
\begin{displaymath}
P_{(i,\ell),(j,\l)}= \frac{\Phi_\ell(i,j)\vhi_\l(j)}{\vhi_\ell(i)}.
\end{displaymath}
\end{theorem}

\proof
We exploit the Markov property of  $X$, rewriting events for $Y$ as events 
for $X$.\\
For every $(j,\l)\in E$ and taking $k\geq 1$ then for any 
$(v_{k},\ell_{k}),\ldots,(v_{1},\ell_{1})\in E \cup \{(m,1)\}$, we denote by  
$A_{k}$ the event~:
\begin{displaymath}
A_{k}=\{Y_{k}=(v_{k},\ell_{k}),\ldots, Y_{1}=(v_{1},\ell_{1})\}.
\end{displaymath}
We have to check that
\begin{displaymath}
\p(Y_{k+1}=(j,\l)|A_{k})=
\p(Y_{2}=(j,\l)|Y_{1}=(v_k,\ell_k)).
\end{displaymath}
First, we observe that
$$A_1 = \{Y_{1}=(v_{1},\ell_{1})\} 
= \{X_1=v_1 < \cdots < X_{\ell_1} > X_{\ell_1 + 1}\},$$
and
\begin{eqnarray*}
A_2 & = & \{Y_{2}=(v_{2},\ell_{2}), Y_{1}=(v_{1},\ell_{1})\} \\
& = & \{X_1=v_1 < \cdots < X_{\ell_1} > X_{\ell_1 + 1}=v_2
< \cdots < X_{\ell_1+\ell_2} > X_{\ell_1 + \ell_2 + 1}\} \\
& = & A_1 \cap 
\{X_{\ell_1 + 1}=v_2 < \cdots < X_{\ell_1+\ell_2} > X_{\ell_1 + \ell_2 + 1}\}.
\end{eqnarray*}
By induction, we obtain 
\begin{eqnarray*}
A_{k}=A_{k-1} \cap 
\{X_{\ell(k-1) + 1}=v_k < \cdots < 
X_{\ell(k)} > X_{\ell(k) + 1}\},
\end{eqnarray*}
where $\ell(k) = \ell_1 + \ldots + \ell_{k}$.
Using this remark and the fact that $X$ is a homogeneous Markov chain, we get

$\DS{\p(Y_{k+1}=(j,\l)|A_{k}) = \p(V_{k+1}=j,L_{k+1}=\l|A_{k})}$
\begin{eqnarray*}
& = & \p(X_{\ell(k) + 1}=j < \cdots < X_{\ell(k)+\l} > X_{\ell(k)+\l + 1} |
X_{\ell(k-1) + 1}=v_k < \cdots < X_{\ell(k)} > X_{\ell(k) + 1}, A_{k-1}) \\
& = & \p(X_{\ell(k) + 1}=j < \cdots < X_{\ell(k)+\l} > X_{\ell(k)+\l + 1} |
X_{\ell(k-1) + 1}=v_k < \cdots < X_{\ell(k)} > X_{\ell(k) + 1}) \\
& = & \p(X_{\ell_k + 1}=j < \cdots < X_{\ell_k+\l} > 
X_{\ell_k+\l + 1} | X_{1}=v_k < \cdots < X_{\ell_k} > X_{\ell_k + 1}) \\
& = & \p(V_{2}=j,L_{2}=\l|V_{1}=v_k,L_{1}=\ell_k) \\
& = & \p(Y_{2}=(j,\l)|Y_{1}=(v_k,\ell_k)).
\end{eqnarray*}
We now have to show that 
$$\p(Y_{k+1}=(j,\l)|Y_{k}=(v_k,\ell_k)) = \p(Y_{2}=(j,\l)|Y_{1}=(v_k,\ell_k)).$$
Using the previous result, we have 
\begin{eqnarray*}
\p(Y_{k+1}=(j,\l)|Y_{k}=(v_k,\ell_k)) & = & \frac{\p(Y_{k+1}=(j,\l),Y_{k}=(v_k,\ell_k))}
{\p(Y_{k}=(v_k,\ell_k))} \\ 
& = & \frac{\DS{\sum_{i=1}^{k-1}\sum_{(v_i,\ell_i) \in E} 
\p(Y_{k+1}=(j,\l),Y_{k}=(v_k,\ell_k),A_{k-1})}}
{\DS{\sum_{i=1}^{k-1}\sum_{(v_i,\ell_i) \in E} 
\p(Y_{k}=(v_k,\ell_k),A_{k-1})}} \\
& = & \frac{\DS{\sum_{i=1}^{k-1}\sum_{(v_i,\ell_i) \in E}
\p(Y_{k+1}=(j,\l)|A_{k})\p(A_{k})}}
{\DS{\sum_{i=1}^{k-1}\sum_{(v_i,\ell_i) \in E} \p(A_{k})}} \\
& = & \p(Y_{2}=(j,\l)|Y_{1}=(v_k,\ell_k)).
\end{eqnarray*}
We have shown that $Y$ is a homogeneous Markov chain over its state space.
The entries of matrix $P$ are then given, for every 
$(j,\l) \in E$ and
$(i,\ell) \in E \cup \{(m,1)\}$ by 
\begin{eqnarray*}
P_{(i,\ell),(j,\l)} & = & \p(V_{k+1} = j, L_{k+1} = \l |V_k = i,L_k = \ell) \\
& = & \p(V_{k+1} = j | V_k = i,L_k = \ell)
\p(L_{k+1} = \l |V_{k+1} = j,V_k = i,L_k = \ell) \\
& = & \p(V_{k+1} = j | V_k = i,L_k = \ell)
\p(L_{k+1} = \l |V_{k+1} = j) \\
& = & \frac{\p(V_{k+1} = \l,L_k = \ell | V_k = i)}
{\p(L_k = \ell| V_k = i)}
\vhi_\l(j) \\
& = & \frac{\Phi_\ell(i,j)\vhi_\l(j)}{\vhi_\ell(i)},
\end{eqnarray*}
where the third equality follows from the Markov property.
\cqfd

We give the expressions of $\vhi_\l(j)$ and $\Phi_\ell(i,j)$ for every 
$i,i,\ell \in S$ in
the following lemma.

\begin{lemm} \label{phi-val}
For every $i,j,\ell \in S$, we have
\begin{eqnarray*} 
\Phi_\ell(i,j) & = & \frac{\DS{{m-i \choose \ell-1}}}{(m-1)^{\ell}}
1_{\{m -i \geq \ell-1\}} 
- \frac{\DS{{j-i \choose \ell-1}}}{(m-1)^{\ell}}
1_{\{j -i \geq \ell-1\}}, \\
\psi_\ell(i) & = & \frac{\DS{{m-i \choose \ell-1}}}{(m-1)^{\ell-1}}
1_{\{m -i \geq \ell-1\}}, \\
\vhi_\ell(i) & = & \frac{\DS{{m-i \choose \ell-1}}}{(m-1)^{\ell-1}}
1_{\{m -i \geq \ell-1\}}
-\frac{\DS{{m-i \choose \ell}}}{(m-1)^{\ell}}
1_{\{m -i \geq \ell\}}.
\end{eqnarray*}       
\end{lemm}

\proof
For every $i,j,\ell \in S$, it is
easily checked that $\Phi_\ell(i,j) = 0$ if $m < i+\ell-1$. 
If
$m \geq i+\ell-1$, we have
\begin{eqnarray}
\Phi_\ell(i,j) & = & \p(V_{2} = j,L_1 = \ell | V_1 = i) \nonumber \\
& = & \p(i < X_2 < \ldots < X_\ell > X_{\ell + 1} = j | X_1 = i) \nonumber \\
& = & \p(i < X_2 < \ldots < X_\ell, X_{\ell + 1} = j | X_1 = i) \nonumber \\
&   &-\p(i < X_2 < \ldots < X_\ell < X_{\ell + 1} = j | X_1 = i)
1_{\{j > i +\ell-1\}}. \label{aa}
\end{eqnarray}
We introduce the sets $G_1(i,j,\ell,m)$, $G_2(i,j,\ell,m)$, $G(i,\ell,m)$ 
and $H(\ell,m)$ defined by
$$G_1(i,j,\ell,m) = \{(x_2,\ldots,x_{\ell+1}) \in 
\{i+1, \ldots,m\}^{\ell} \, ; \,
     x_2 < \cdots < x_\ell \neq x_{\ell+1}=j\},$$
$$G_2(i,j,\ell,m) = \{(x_2,\ldots,x_{\ell+1}) \in 
\{i+1, \ldots,m\}^{\ell} \, ; \,
     x_2 < \cdots < x_\ell =  x_{\ell+1}=j\},$$
$$G(i,\ell,m) = \{(x_2,\ldots,x_{\ell}) \in 
\{i+1, \ldots,m\}^{\ell - 1} \, ; \,
     x_2 < \cdots < x_\ell\},$$
$$H(\ell,m) = \{(x_2,\ldots,x_{\ell+1}) \in 
\{1, \ldots,m\}^{\ell} \, ; \,
     i \neq x_2 \neq \cdots \neq x_{\ell + 1}\}.$$
It is well-known, see for instance \cite{Foata96}, that
$$|G(i,\ell,m)| = {m-i \choose \ell-1}.$$
Since $|G_2(i,j,\ell,m)| = |G(i,\ell-1,j-1)|$, the first term in (\ref{aa}) 
can be written as 
\begin{eqnarray*}
\p(i < X_2 < \ldots < X_\ell, X_{\ell + 1} = j | X_1 = i) 
& = & \frac{\left|G_1(i,j,\ell,m)\right|}{\left|H(\ell,m)\right|} \\
& = & \frac{\left|G(i,\ell,m)\right|-\left|G_2(i,j,\ell,m)\right|}
     {\left|H(\ell,m)\right|} \\
& = & \frac{\left|G(i,\ell,m)\right|-\left|G(i,\ell-1,j-1)\right|}
     {\left|H(\ell,m)\right|} \\
& = &\frac{\DS{{m-i \choose \ell-1}} - \DS{{j-i-1 \choose \ell-2}}
          1_{\{j-i \geq \ell-1\}}}{(m-1)^{\ell}},
\end{eqnarray*}
The second term is given, for $j > i +\ell-1$, by
$$\p(i < X_2 < \ldots < X_\ell < X_{\ell + 1} = j | X_1 = i)
= \frac{\left|G(i,\ell,j-1)\right|}{\left|H(\ell,m)\right|} 
= \frac{\DS{{j-i-1 \choose \ell-1}}}{(m-1)^{\ell}}.$$
Adding these two terms, we get
\begin{eqnarray*}
\Phi_\ell(i,j) & = & \frac{\DS{{m-i \choose \ell-1}}1_{\{m-i \geq \ell-1\}}
-\DS{{j-i-1 \choose \ell-2}}
1_{\{j-i \geq \ell-1\}} - 
\DS{{j-i-1 \choose \ell-1}}1_{\{j-i \geq \ell\}}}
{(m-1)^{\ell}} \\
& = & \frac{\DS{{m-i \choose \ell-1}}1_{\{m-i \geq \ell-1\}}
-\DS{{j-i \choose \ell-1}}
1_{\{j-i \geq \ell-1\}}}{(m-1)^{\ell}},
\end{eqnarray*}
which completes the proof of the first relation.

The second relation follows from expression (\ref{psiil}) by writing
\begin{eqnarray*}
\psi_\ell(i) & = & \p(L_1 \geq \ell | V_1 = i) \\
& = & \p(i < X_2 < \ldots < X_\ell | X_1 = i)
1_{\{m-i \geq \ell-1\}} \\
& = & \frac{\left|G(i,\ell,m)\right|}{\left|H(\ell-1,m)\right|} \\
& = & \frac{\DS{{m-i \choose \ell-1}}}{(m-1)^{\ell-1}}1_{\{m-i \geq \ell-1\}}.
\end{eqnarray*}
The third relation follows from definition (\ref{fiil}) by writing
$\vhi_\ell(i) = \psi_\ell(i) - \psi_{\ell+1}(i)$.
\cqfd
Note that the matrix $\Phi$ defined by
$$\Phi = \sum_{\ell=1}^m \Phi_\ell$$
is obviously a stochastic matrix, which means that, for every
$i = 1, \ldots,m$, we have
$$\sum_{\ell=1}^m \vhi_\ell(i) = 1.$$
$$\sum_{\ell=1}^m \sum_{j = 1}^m \Phi_\ell(i,j) =
\sum_{\ell=1}^m \vhi_\ell(i) = \psi_(i) = 1.$$

\section{Hitting times and maximal ascending run}
\label{sec:hitting}
For every $r = 1,\ldots,m$, we denote by $T_r$ the hitting time of an 
ascending run of
length at least equal to $r$. More formally, we have
\begin{displaymath}
T_r = \inf\{k \geq r\,;\, X_{k-r+1} < \cdots < X_{k}\}.
\end{displaymath}
It is easy to check that we have $T_1 = 1$ and $T_r \geq r$.
The distribution of $T_r$ is given by the following theorem.

\begin{theorem} \label{loideTr}
For $2 \leq r \leq m$, we have
\begin{equation} \label{loi}
\p(T_r \leq n | V_1 = i) = \left\{\begin{array}{l}
                           0 \mbox{ ~if~ } 1 \leq n \leq r-1 \\\\
                    \DS{\psi_r(i)
             + \sum_{\ell = 1}^{r-1}\sum_{j=1}^{m} \Phi_\ell(i,j)
               \p(T_r \leq n-\ell | V_1 = j)} \mbox{ ~if~ } n \geq r.
\end{array}\right.
\end{equation}
\end{theorem}

\proof
Since $T_r \geq r$, we have, for $1\leq n\leq r-1$,
\begin{displaymath}
\p(T_r \leq n | V_1 = i) = 0
\end{displaymath}
Let us assume from now that $n \geq r$. 
Since  $L_1 \geq r$ implies that $T_r = r$, we get
\begin{equation} \label{m1}
\p(T_r \leq n, L_1 \geq r | V_1 = i) = \p(L_1 \geq r | V_1 = i)
= \psi_r(i).
\end{equation}
We introduce the random variable $T^{(p)}_r$ defined by hitting time of an 
ascending run length at least equal to $r$ when counting from position $p$. 
Thus we have
\begin{displaymath}
T^{(p)}_r = \inf\{k \geq r \,;\, X_{p+k-r} < \cdots < X_{p+k-1}\}.
\end{displaymath}
We then have $T_r = T^{(1)}_r$. Moreover, $L_1 = \ell < r$ implies that 
$T_r = T^{(L_1+1)}_r + \ell$, which leads to
\begin{align}
\p(T_r \leq n, L_1 < r | V_1 = i) 
& =  \sum_{\ell = 1}^{r-1}\p(T_r \leq n, L_1 = \ell | V_1 = i) \nonumber \\
& =  \sum_{\ell = 1}^{r-1}
 \p(T^{(L_1+1)}_r \leq n-\ell, L_1 = \ell | V_1 = i) \nonumber \\
& =  \sum_{\ell = 1}^{r-1}\sum_{j=1}^{m} 
 \p(T^{(L_1+1)}_r \leq n-\ell, V_2 = j, L_1 = \ell | V_1 = i) \nonumber \\
& = \sum_{\ell = 1}^{r-1}\sum_{j=1}^{m} \Phi_\ell(i,j)\,
 \p(T^{(L_1+1)}_r \leq n-\ell | V_2 = j, L_1 = \ell, V_1 = i) \nonumber \\
& =  \sum_{\ell = 1}^{r-1}\sum_{j=1}^{m} \Phi_\ell(i,j) \,
 \p(T^{(L_1+1)}_r \leq n-\ell | V_2 = j) \nonumber \\
& =  \sum_{\ell = 1}^{r-1}\sum_{j=1}^{m} \Phi_\ell(i,j)\,
     \p(T_r \leq n-\ell | V_1 = j), \label{m2} 
\end{align}
where the fifth equality follows from the Markov property and the last one
from the homogeneity of $Y$.
Putting together relations (\ref{m1}) and (\ref{m2}), we obtain
\begin{displaymath}
\p(T_r \leq n | V_1 = i) = \psi_r(i)
                  + \sum_{\ell = 1}^{r-1}\sum_{j=1}^{m} \Phi_\ell(i,j)
                              \p(T_r \leq n-\ell | V_1 = j).
\end{displaymath}
\cqfd

For every $n \geq 1$, we define $M_n$ as the maximal ascending run length
over the $n$ first values $X_1, \ldots, X_n$. We have
$1 \leq M_n \leq m \wedge n$ and
\begin{displaymath}M_n \geq r \Longleftrightarrow T_r \leq n,\end{displaymath}
which implies
\begin{displaymath}\e(M_n) = \sum_{r = 1}^{m \wedge n} \p(M_n \geq r) = 
\sum_{r = 1}^{m \wedge n} \p(T_r \leq n) = 
\frac{1}{m}\sum_{r = 1}^{m \wedge n} \sum_{i=1}^m \p(T_r \leq n|V_1=i).
\end{displaymath}

\section{Algorithm}
\label{sec:algorithm}

For $r = 1, \ldots,m$, we denote by $\psi_r$ the column vector of dimension 
$m$ which $i$th entry is $\psi_r(i)$.
For $r = 1, \ldots,m$, $n \geq 1$ and $h = 1,\ldots,n$, we denote by $W_{r,h}$ 
the column vector of dimension $m$ which $i$th 
entry is defined by
$$W_{h,r}(i) = \p(T_r \leq h|V_1=i) = \p(M_h \geq r|V_1=i),$$
and we denote by $\nbu$ the column vector of dimension $m$ with all 
entries equal 
to $1$. 
An algorithm for the computation of the distribution and the expectation
of $M_n$ is given in Table 1.

\begin{tabbing}
  {\bf input :}  $m$, $n$ \\
  {\bf output :} $\e(M_h)$ for $h=1,\ldots,n$. \\
  {\bf for} \= $\ell = 1$ {\bf to} $m$ {\bf do} 
  Compute the matrix $\Phi_\ell$ 
  {\bf endfor} \\
  {\bf for} \= $r = 1$ {\bf to} $m$ {\bf do} 
  Compute the column vectors $\psi_r$ 
  {\bf endfor} \\ 
  {\bf for} \= $h = 1$ {\bf to} $n$ {\bf do} 
  $W_{h,1} = \nbu$ 
  {\bf endfor} \\
  {\bf for} \= $r = 2$ {\bf to} $m \wedge n$ {\bf do} \\
  \> {\bf for} \= $h = 1$ {\bf to} $r-1$ {\bf do} 
     $W_{h,r} = 0$ {\bf endfor} \\
  \> \> ttttt\=ttttt\=ttttt\=ttttt \kill
  \> {\bf for} \= $h = r$ {\bf to} $n$ {\bf do} 
               $\DS{W_{h,r} = \psi_r + \sum_{\ell = 1}^{r-1} 
                                       \Phi_\ell W_{h - \ell,r}}$ 
               {\bf endfor} \\
  {\bf endfor} \\
  {\bf for} \= $h = 1$ {\bf to} $n$ {\bf do}
           $\DS{\e(M_h) = \frac{1}{m}\sum_{r = 1}^{m \wedge h} \nbu^t W_{h,r}}$
  {\bf endfor} \\
\end{tabbing}
\begin{center}
{Table 1: Algorithm for the distribution and expectation computation of $M_n$.}
\end{center}

\section{\hspace{-0.4cm}Application to wireless networks : fast self-organization}
\label{sec:appli}

Our analysis has important implications in forecast large-scale
wireless networks. In those networks, the number of machines involved
and the likeliness of fault occurrences prevents any centralized
planification. Instead, distributed self-organization must be designed
to enable proper functioning of the network. A useful technique to
provide self-organization is
\emph{self-stabilization}~\cite{D74j,D00b}. Self-stabilization is a
versatile technique that can make a wireless network withstand any
kind of fault and reconfiguration.

A common drawback with self-stabilizing protocols is that they were
not designed to handle properly large-scale networks, as the stabilizing
time (the maximum amount of time needed to recover from any possible
disaster) could be related to the actual size of the network. In many
cases, this high complexity was due to the fact that network-wide
unique identifiers are used to arbitrate symmetric
situations~\cite{T07c}. However, there exists a number of problems
appearing in wireless networks that need only locally unique
identifiers.

Modeling the network as a graph where nodes represent wireless
entities and where edges represent the ability to communicate between
two entities (because each is within the transmission range of the
other), a local coloring of the nodes at distance $d$ (\emph{i.e.}
having two nodes at distance $d$ or less assigned a
distinct color) can be enough to solve a wide range of problems. For
example, local coloring at distance $3$ can be used to assign TDMA
time slots in an adaptive manner~\cite{HT04c}, and local coloring at
distance $2$ has successively been used to self-organize a wireless
network into more manageable clusters~\cite{WWAN05}.

In the performance analysis of both schemes, it appears that the overall
stabilization time is balanced by a tradeoff between the coloring time itself
and the stabilization time of the protocol using the coloring (denoted in the
following as the \emph{client} protocol). In both cases
(TDMA assignment and clustering), the stabilization time of the client
protocol is related to the
height of the directed acyclic graph induced by the colors. This DAG is
obtained by orienting an edge from the node with the highest color to
the neighbor with the lowest color. As a result, the overall height of
this DAG is equal to the longest strictly ascending chain of colors across
neighboring nodes.
Of course, a larger set of colors leads to a shorter stabilization
time for the coloring (due to the higher chance of picking a fresh color),
but yields to a potential higher DAG, that could delay the stabilization time
of the client protocol.

In~\cite{ICPADS06}, the stabilization time of the coloring protocol was
theoretically analyzed while the stabilization time of a particular
client protocol (the clustering scheme of~\cite{WWAN05}) was only
studied by simulation.
The analysis performed in this paper gives a theoretical upper bound
on the stabilization time of all client protocols that use a coloring
scheme as an underlying basis. Together with the results of~\cite{ICPADS06},
our study constitutes a comprehensive analysis of the overall stabilization
time of a class of self-stabilizing protocols used for the self-organization
of wireless sensor networks. In the remaining of the section, we provide
quantitative results regarding the relative importance of the number of used
colors with respect to other network parameters.

\begin{figure}
\centering
\includegraphics[width=10cm,angle=-90]{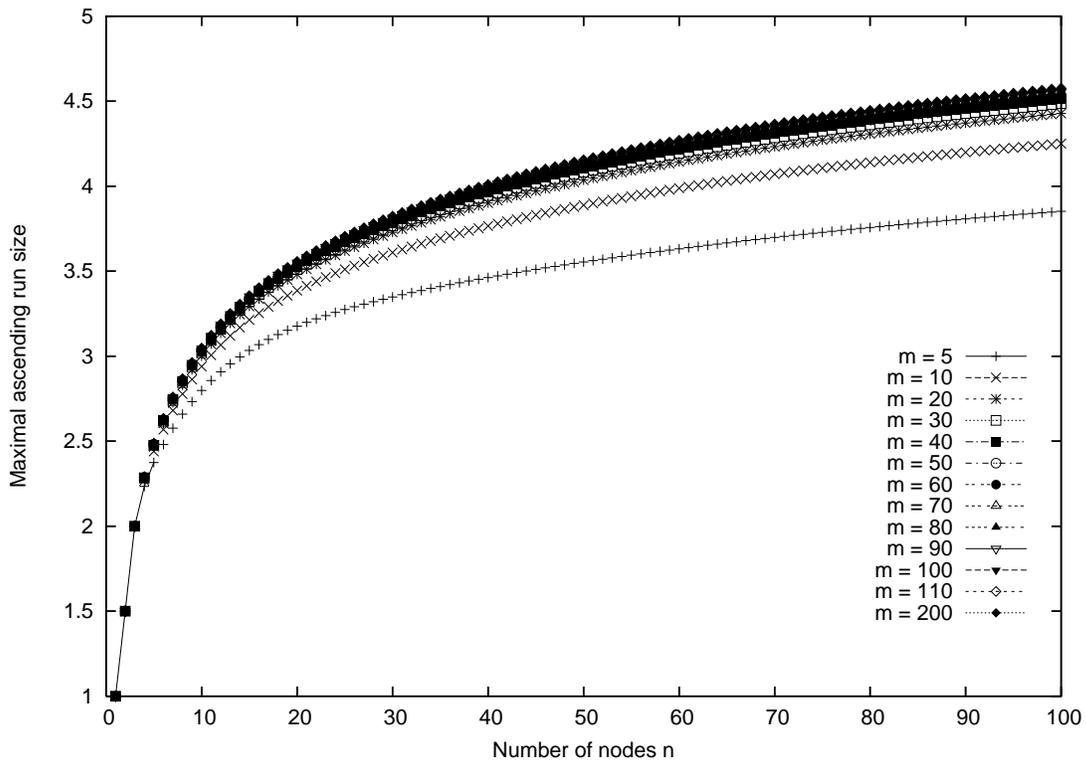}
\caption{Expected length of the maximal ascending run as a function of the number of nodes.}
\label{fig:lgfctn}
\end{figure}

Figure~\ref{fig:lgfctn} shows the expected length of the maximal ascending run
over a $n$-node chain for different values of $m$. 

Results show several interesting behaviors. Indeed, self-organization protocols 
relying on a coloring process achieve better stabilization time when the 
expected length of maximal ascending run is short but a coloring process stabilizes 
faster when the number of colors is high \cite{ICPADS06}. 

Figure~\ref{fig:lgfctn} clearly shows that even if the number of colors 
is high compared to $n$ ($n<<m$), the expected length of maximal ascending run 
remains short, which is a great advantage. 
Moreover, even if the number of nodes increases, the expected length 
of the maximal 
ascending run remains short and increases very slowly. This observation demonstrates
the scalability properties of a protocol relying on a local coloring process since its 
stabilization time is directly linked to the length of this 
ascending run~\cite{ICPADS06}. 

\begin{figure}
\centering
\includegraphics[width=10cm,angle=-90]{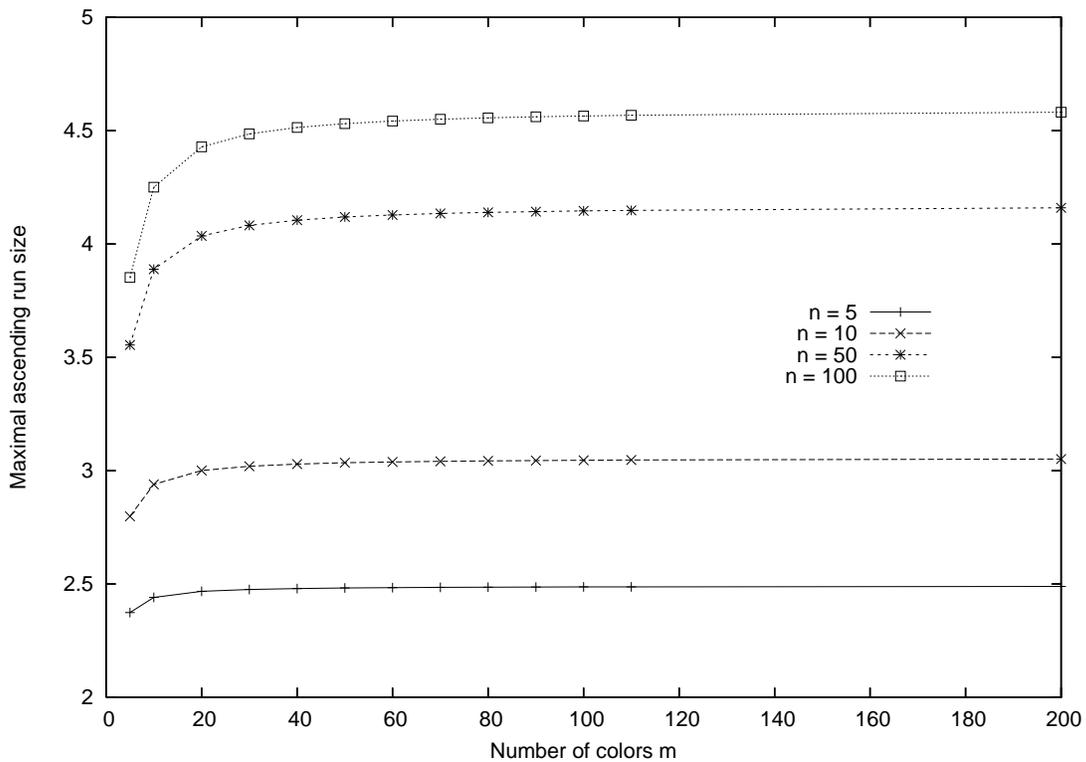}
\caption{Expected length of the maximal ascending run as a function of the number of colors.}
\label{fig:lgfctm}
\end{figure}

Figure~\ref{fig:lgfctm} shows the expected length of maximal ascending run over 
a $n$-node chain for different values of $n$. 

Results shows that for a fixed number of nodes $n$, the expected length of the maximal 
ascending run converges to a finite value, depending of $n$. This implies that using a large number of colors does not impact the stabilization time of the client algorithm. 

\bibliographystyle{plain}
\bibliography{runs}

\end{document}